\documentclass[12pt]{article}
\usepackage{graphicx}
\def\nn{\noindent}
\def\Re{{\cal R \mskip-4mu \lower.1ex \hbox{\it e}\,}}
\def\Im{{\cal I \mskip-5mu \lower.1ex \hbox{\it m}\,}}
\def\ie{{\it i.e.}}
\def\eg{{\it e.g.}}

\def\sub#1{_{\lower.25ex\hbox{$\scriptstyle#1$}}}
\def\tev{\,{\ifmmode\mathrm {TeV}\else TeV\fi}}
\def\gev{\,{\ifmmode\mathrm {GeV}\else GeV\fi}}
\def\mev{\,{\ifmmode\mathrm {MeV}\else MeV\fi}}
\def\mpl{\ifmmode M_{pl}\else $M_{pl}$\fi}
\def\to{\rightarrow}

\def\subw{_{\rm w}}
\def\mh{\ifmmode m\sbl H \else $m\sbl H$\fi}
\def\mch{\ifmmode m_{H^\pm} \else $m_{H^\pm}$\fi}
\def\mt{\ifmmode m_t\else $m_t$\fi}
\def\mc{\ifmmode m_c\else $m_c$\fi}
\def\mz{\ifmmode M_Z\else $M_Z$\fi}
\def\mw{\ifmmode M_W\else $M_W$\fi}
\def\mws{\ifmmode M_W^2 \else $M_W^2$\fi}
\def\mhs{\ifmmode m_H^2 \else $m_H^2$\fi}   
\def\mzs{\ifmmode M_Z^2 \else $M_Z^2$\fi}
\def\mts{\ifmmode m_t^2 \else $m_t^2$\fi}
\def\mcs{\ifmmode m_c^2 \else $m_c^2$\fi}
\def\mchs{\ifmmode m_{H^\pm}^2 \else $m_{H^\pm}^2$\fi}
\def\ztwo{\ifmmode Z_2\else $Z_2$\fi}
\def\zone{\ifmmode Z_1\else $Z_1$\fi}
\def\mtwo{\ifmmode M_2\else $M_2$\fi}
\def\mone{\ifmmode M_1\else $M_1$\fi}
\def\tb{\ifmmode \tan\beta \else $\tan\beta$\fi}
\def\xw{\ifmmode x\subw\else $x\subw$\fi}
\def\ch{\ifmmode H^\pm \else $H^\pm$\fi}
\def\lum{\ifmmode {\cal L}\else ${\cal L}$\fi}
\def\inpb{\,{\ifmmode {\mathrm {pb}}^{-1}\else ${\mathrm {pb}}^{-1}$\fi}}
\def\infb{\,{\ifmmode {\mathrm {fb}}^{-1}\else ${\mathrm {fb}}^{-1}$\fi}}
\def\epem{\ifmmode e^+e^-\else $e^+e^-$\fi}
\def\ppb{\ifmmode \bar pp\else $\bar pp$\fi}
\def\bsg{\ifmmode B\to X_s\gamma\else $B\to X_s\gamma$\fi}
\def\bsll{\ifmmode B\to X_s\ell^+\ell^-\else $B\to X_s\ell^+\ell^-$\fi}
\def\bstt{\ifmmode B\to X_s\tau^+\tau^-\else $B\to X_s\tau^+\tau^-$\fi}
\def\lamt{\ifmmode \tilde\lambda\else $\tilde\lambda$\fi}
\def\shat{\ifmmode \hat s\else $\hat s$\fi}
\def\that{\ifmmode \hat t\else $\hat t$\fi}
\def\uhat{\ifmmode \hat u\else $\hat u$\fi}

\newskip\zatskip \zatskip=0pt plus0pt minus0pt
\def\matth{\mathsurround=0pt}

\def\atversim#1#2{\lower0.7ex\vbox{\baselineskip\zatskip\lineskip\zatskip
  \lineskiplimit 0pt\ialign{$\matth#1\hfil##\hfil$\crcr#2\crcr\sim\crcr}}}

\renewcommand{\thefootnote}{\fnsymbol{footnote}}

\begin{document} \begin{titlepage} 
\rightline{\vbox{\halign{&#\hfil\cr
&SLAC-PUB-10001\cr
&June 2003\cr}}}
\begin{center}

{\Large\bf More Transverse Polarization Signatures of Extra Dimensions 
at Linear Colliders}
\footnote{Work supported by the Department of 
Energy, Contract DE-AC03-76SF00515}
\medskip

\normalsize 
{\bf \large Thomas G. Rizzo}
\vskip .3cm
Stanford Linear Accelerator Center \\
Stanford University \\
Stanford CA 94309, USA\\
\vskip .2cm

\end{center}

\begin{abstract}
Polarization of both electron and positron beams at a future linear 
collider (LC) allows for the measurement of transverse polarization 
asymmetries. These asymmetries have been shown to be particularly sensitive 
to graviton or other spin-2, $s-$channel exchanges in the process 
$e^+e^- \to f\bar f~(f\neq e)$ which allows for a doubling of the usual 
search reach. A question then arises as to whether other $e^+e^-$ 
processes also show comparable sensitivity. Here we extend our previous 
analysis to the set of final states  $e^+e^-, W^+W^-,~2\gamma$ and $~2Z$ 
as well as to the M{\o}ller scattering process $e^-e^-\to e^-e^-$. We 
demonstrate that these reactions yield transverse polarization asymmetries 
which are somewhat less sensitive to graviton exchange than are those 
obtained in our earlier analysis for $e^+e^-\to f\bar f$.
\end{abstract} 



\renewcommand{\thefootnote}{\arabic{footnote}} \end{titlepage}


\section{Introduction}

The existence of new physics (NP) beyond the Standard Model (SM) at or 
near the TeV scale is anticipated on rather general grounds. Future 
colliders, the LHC and/or a Linear Collider (LC), may be above threshold for 
the production of new particle states, such as SUSY, in which case the NP 
will be observed directly. 
In such a scenario the detailed analysis of the NP will be relatively 
straightforward though it may take some time and the combined data from 
both colliders to accomplish. 
Alternatively, experiments may uncover new reactions with 
small rates which are forbidden in the SM, thus pointing at NP; it may be very 
difficult in such cases to access the details of the 
underlying theory without observing the direct production of the new particles 
inducing these processes. The appearance of 
NP may, however,  be even more subtle than either of these scenarios. We can 
imagine that collider data begin to show small deviations from the SM 
predictions for various observables, \eg, cross sections and asymmetries, 
which grow with increasing energy. A set of such observations signals the 
existence of NP beyond the collider's kinematic reach which is manifesting 
itself in the form of higher dimensional operators, \ie, generalized 
contact interactions. In the more complete theory at higher energies 
these operators 
arise from the exchanges of new particles, which are too massive to be 
directly produced at the collider. These particles may occur with 
different spins and in various channels depending upon the particular theory. 
The literature contains a rather long 
list of potential NP scenarios of this type that lead to contact interactions 
once the heavy fields are integrated out: a $Z'$ {\cite {e6,zp}}, scalar 
or vector leptoquarks{\cite {e6,lq}, 
$R$-parity violating sneutrino($\tilde \nu$) exchange{\cite {rp}}, scalar or 
vector bileptons{\cite {bl}}, graviton Kaluza-Klein(KK) 
towers{\cite {ed,dhr}} in extra dimensional models{\cite {add,rs}}, 
gauge boson KK towers{\cite {ed2,dhr}}, and even string 
excitations{\cite {se}}. It is this type of observation of NP that we will 
discuss in this paper. 

It is clear that it is important to develop techniques that will help in 
distinguishing among the many possible sources of contact interactions. 
In a previous paper{\cite {tp1}} we have considered one such possibility 
at the LC: transverse polarization (TP) and the associated 
asymmetries{\cite {tp}} that can be formed through its use. Provided both 
the $e^-$ and $e^+$ beams can be initially longitudinally polarized, spin 
rotators can then used to convert the longitudinal polarization to transverse 
polarization with near to $100\%$ efficiencies. As the reader may recall and 
as we will see below, double beam polarization 
is necessary to generate the TP asymmetries. While historically 
the possible use of TP as a tool for new physics searches  
has not gotten much attention{\cite {tp}}, our earlier analysis of TP 
asymmetries for the process $e^+e^-\to f\bar f~(f\neq e)$ found them 
to be a unique probe for the $s$-channel exchange of spin-2 fields, 
especially when we sum over all of the accessible final states, $f$.  
Currently, we associate such exchanges with the Kaluza-Klein graviton 
towers of the Arkani-Hamed, Dimopoulos and Dvali(ADD){\cite {add}} or 
Randall-Sundrum(RS){\cite {rs}} scenarios. 

The purpose of the present paper is to extend the previous analysis to 
other final states which are also accessible in $e^+e^-$ collisions: $e^+e^-, 
~W^+W^-,~2\gamma$ and $2Z$; for completeness we also 
include the M{\o}ller scattering process $e^-e^-\to e^-e^-$. Here we wish 
to explore whether any of these final states lead to TP asymmetries 
which are also particularly sensitive to spin-2/graviton 
exchanges.  Unfortunately, 
we will show that this is not the case. Though the discovery reach for each 
of these processes is significant, it is always somewhat less than that 
found for the sum over the $f\bar f$ final 
states obtained in our earlier analysis.  

The organization of this paper is as follows. After our introduction, we will 
provide a brief overview and review of TP and the associated asymmetries in 
$e^+e^-$ collisions generalizing the formalism from our previous discussion 
of the $f\bar f$ final states to accommodate those that are considered here. 
In section 3 we analyze in turn the TP asymmetries for each of the final 
states $e^+e^-, W^+W^-, 2\gamma$ and $2Z$ as well as $e^-e^- \to e^-e^-$ 
in both the SM and the ADD model. In 
particular we show how the SM predictions for TP asymmetries are modified 
by the presence of spin-2 graviton exchange. Subsequently, the search reaches 
arising from the deviations in the TP asymmetries for the ADD model are 
obtained for each of these final states. The use of these final states for 
uniquely identifying graviton exchange is also analyzed. 
A discussion and our conclusions can be found in section 4.

\section{Transverse Polarization Asymmetries: Background}

Much of the formalism regarding TP and the associated asymmetries can be 
found in our earlier work{\cite {tp1}}. Here we will provide only a quick 
overview and the necessary background required to follow the analysis 
we present below. 

Consider the set of processes  
$e^+e^- \to X \bar X$ with the both electron and positron beams polarized. 
Taking the initial $e^{\pm}$ beam momenta along the $\mp z-$axis
we denote the longitudinal and transverse polarizations of the $e^-(e^+)$ 
by their cartesian components $P_{x,y,z}(\bar P_{x,y,z})$. 
For the moment we allow these two 
polarization vectors to be arbitrarily oriented. To proceed, we will follow a 
modified version of the notation used by Hikasa{\cite {tp}} and denote 
the corresponding helicity amplitudes for this process by 
$T_{h\bar h}^{h'\bar h'}$ where $h(\bar h)$ represent the 
$\pm$ helicity of the initial $e^-(e^+)$ and $h'(\bar h')$ is the 
corresponding helicity of $X(\bar X)$. Considering the cases of interest,   
$X=f,e,\gamma,W,Z$, we find that many of the products of these amplitudes 
cancel when summed over final helicities even when graviton exchange is 
included. For example, in the case of fermion pairs in the final state,  
we obtain
\begin{eqnarray}
\sum_{ij} ~T_{+-}^{*~ij}~T_{\pm\pm}^{~ij} &=& 0\, \nonumber \\ 
\sum_{ij} ~T_{\pm\pm}^{*~ij}~T_{-+}^{~ij} &=& 0\, \nonumber \\ 
\sum_{ij} ~T_{++}^{*~ij}~T_{--}^{~ij} &=& 0\,. 
\end{eqnarray}
Similar equalities hold for the case of gauge bosons in the final state and, 
due to crossing symmetry,   
for the M{\o}ller scattering process $e^-e^-\to e^-e^-$. 
When these conditions hold, terms in the spin-averaged matrix element 
proportional to either $P_{x,y}$, $\bar P_{x,y}$ individually or the products 
$P_z \bar P_{x,y}$ and $P_{x,y} \bar P_z$ are all seen to vanish. In this case 
the spin-averaged matrix element for this class of processes can 
be symbolically written as
\begin{eqnarray}
|\bar {\cal M}|^2&=&{1\over {4}}\Big[(|T_{+-}|^2+|T_{-+}|^2+|T_{++}|^2
+|T_{--}|^2) 
+P_z(|T_{+-}|^2-|T_{-+}|^2+|T_{++}|^2-|T_{--}|^2)\, \nonumber \\ 
&+&\bar {P_z}(|T_{+-}|^2-|T_{-+}|^2-|T_{++}|^2+|T_{--}|^2)\, \nonumber \\
&+&P_z\bar {P_z}
(|T_{+-}|^2+|T_{-+}|^2-|T_{++}|^2-|T_{--}|^2)\, \nonumber \\
&+&2(P_x\bar {P_x}-P_y\bar {P_y})~Re(T^*_{+-}T_{-+})\cos 2\phi
+2(P_x\bar {P_y}+P_y\bar {P_x})~Im(T^*_{+-}T_{-+})\sin 2\phi\Big]\,,
\end{eqnarray}
where $\phi$ is the azimuthal angle and the implied summations over the final 
state helicities in each product of amplitudes, ${ij}$, are suppressed. 
As in the work of Hikasa{\cite {tp}}, the 
helicity amplitudes in the expression above are now defined with the 
angle $\phi$ set to zero. 
Note that the $T_{\pm\pm}$ amplitudes only appear quadratically. 
We observe from this expression 
the important fact that the $\phi$-dependent pieces are {\it only} 
accessible if both beams are simultaneously 
transversely polarized. Thus we are reminded that 
to have azimuthal transverse polarization asymmetries at a LC 
we {\it must} begin with both beams longitudinally polarized and employ spin 
rotators; this differs from the case of the usual left-right (longitudinal) 
polarization asymmetry, $A_{LR}$, which requires only single $e^-$ beam 
polarization, \ie, the term above linear in $P_z$.

In what follows  we will for simplicity assume in our analysis that we are 
in an energy regime where the 
effects of the finite width of the $Z$ can be neglected; in addition 
we will assume that we can also safely neglect (as is usual) the small 
imaginary contributions to the amplitudes arising from 
graviton exchange{\cite {stuff}} in the ADD model so that the `$Im$' 
terms in the expression above can be dropped. In forming the TP asymmetries 
we will limit ourselves to 
the case where the beams are purely transversely polarized with the 
directions of polarization vectors being back-to-back. We can define $\phi$ 
to be the angle between the $e^\pm$ polarization directions and the plane 
of the momenta of the outgoing $X\bar X$ particles in the final state. 

Given the squared matrix element we can now form as before the differential 
azimuthal asymmetry distribution which we symbolically define by 
\begin{equation}
{1\over {N}} {dA\over {dz}}=\Bigg[{{\int_+ {d\sigma \over {dz d\phi}}-\int_- 
{d\sigma \over {dz d\phi}}}\over {\int d\sigma}}\Bigg]\,,
\end{equation}
where $\int_{\pm}$ are integrations over regions where $\cos 2\phi$ takes on 
$\pm$ values; integration over the full ranges of $z$ and $\phi$ occurs 
in the denominator, except for possible acceptance cuts or cuts employed to 
remove QED $t-$ and $u-$ channel poles. 
(Here, $z=\cos \theta$, is the usual scattering angle.) 
Note that the presence or absence of the `$Im$' terms 
proportional to $\sin 2\phi$ 
do not influence the value of this asymmetry since they cancel in both the 
numerator and denominator. 
It is important to recall that we expect this differential 
asymmetry to take on rather small numerical 
values since it is normalized to the total cross section and {\it not} to the 
differential cross section at the same value of $z$ as is usually done.  
This isolates the important angular behavior in the numerator of the 
TP asymmetry where it can be much more easily studied.
In the case of $e^+e^-\to f\bar f~~(f\neq e)$, we found that this asymmetry 
was proportional to $1-z^2$ in the SM as well as in most of the SM 
extensions discussed above. Only in the case of spin-2/graviton  exchange 
was there a significant distortion of the angular dependence of this 
asymmetry thus leading to a unique signature for this special kind of NP.

\section{Analysis}

In order to begin our analysis we need the full set of helicity amplitudes 
for the above processes including the contributions from spin-2 graviton 
exchange. These can be found in the literature, \eg, 
{\cite {agashe,tp}} and through the use of crossing symmetry. In the 
ADD scenario, which we will concentrate on in what 
follows, the relative contribution of the spin-2 graviton to these 
amplitudes always appears with a suppression factor, $f_g$; employing 
the convention of Hewett{\cite {ed}}, this is given by   
\begin{equation}
f_g={\lambda s^2\over {4\pi \alpha M_H^4}}\,. 
\end{equation}
where $M_H$ represents the cutoff scale in the Kaluza-Klein (KK) 
graviton tower sum 
and $\lambda=\pm 1$. This factor clearly shows the dimension-8 origin of 
the gravitational $\sim T_{\mu\nu}T^{\mu\nu}$ interaction induced 
in the ADD model after summing over the KK tower. 
In the RS model, to which our analysis is easily generalized, the 
corresponding expression can be obtained through the replacement
\begin{equation}
{\lambda \over {M_H^4}} \to {-1\over {8\Lambda_\pi^2}}\sum_n 
{1\over {s-m_n^2+im_n\Gamma_n}}\,. 
\end{equation}
where $\Lambda_\pi$ is of order a few TeV and $m_n(\Gamma_n)$ are the 
masses(widths) of the TeV scale graviton KK excitations. 

With the amplitudes in hand we can directly proceed to calculate the 
azimuthal polarization asymmetries. For numerical purposes we will assume 
that $P_{e^-}=80\%$ and $P_{e^+}=60\%$ as in our previous analysis. 
We remind the reader that 
by combining the data for the various $f\bar f$ final states, 
$f=\mu,\tau,c,b,t$, search reaches, \ie, 
$95\%$ CL bounds, as large as $M_H \sim 20 \sqrt s$ were obtained in our 
earlier work using {\it only} the TP asymmetry in the fits. Similarly, 
the identification (ID) reach, the value of $M_H$ 
for which a $5\sigma$ signal for spin-2 exchange is observed, was found to 
be $M_H \sim 10\sqrt s$. These values are roughly a factor of 2 greater 
than those obtained from more conventional analyses{\cite {cont}}. It is to 
these values that we must compare the results we obtain below. 

\vspace*{-0.5cm}
\nn
\begin{figure}[htbp]
\centerline{
\includegraphics[width=8cm,angle=90]{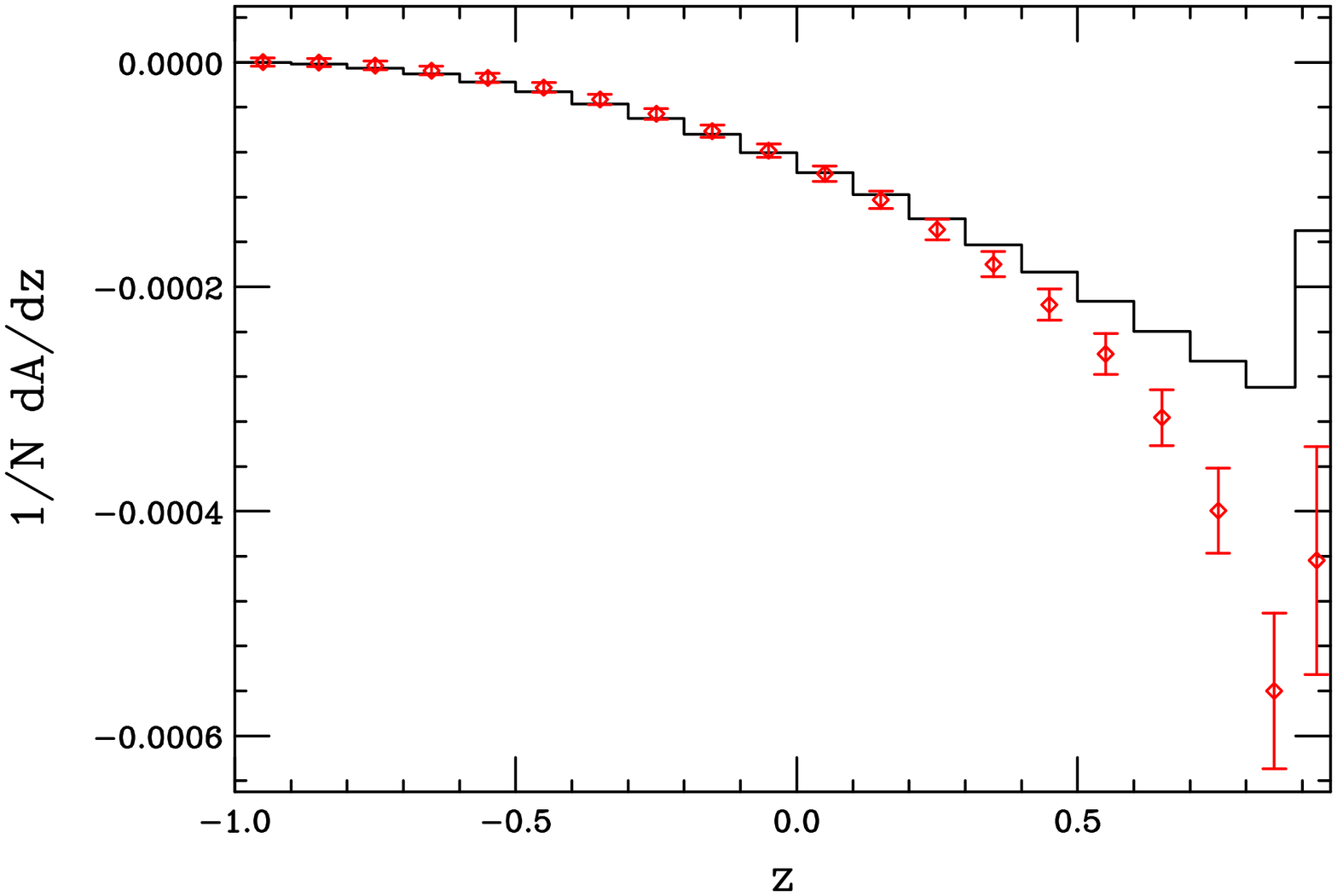}}
\vspace*{0.45cm}
\centerline{
\includegraphics[width=8cm,angle=90]{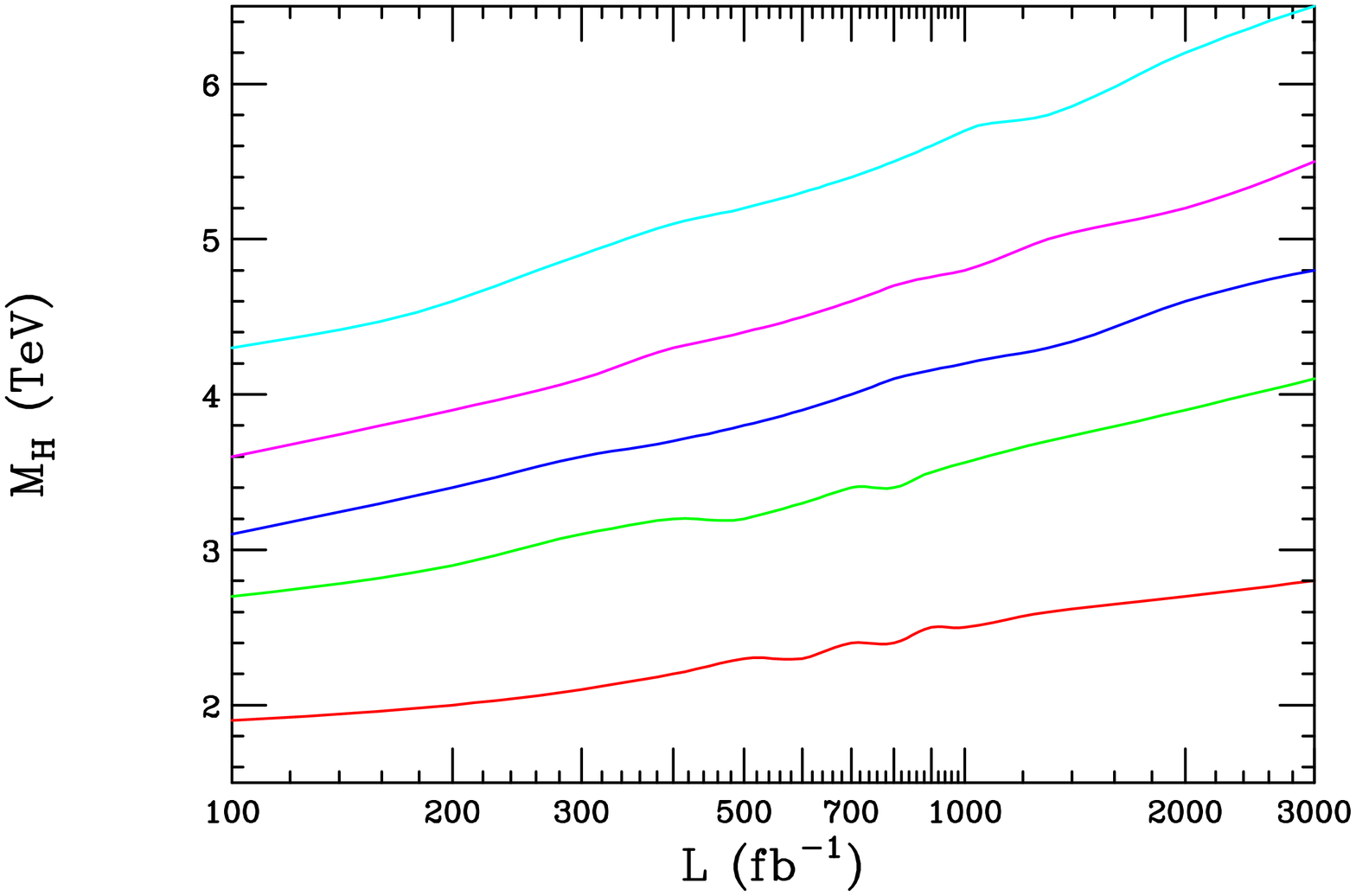}}
\vspace*{0.25cm}
\caption{(Top)Differential transverse polarization azimuthal 
asymmetry distribution for Bhabha scattering, 
$e^+e^-\to e^+e^-$, at a 500 GeV LC assuming a luminosity of 500 $fb^{-1}$. 
The histogram is the SM prediction while the data points are for the ADD 
model assuming $M_H=1.5$ TeV. The errors shown are the quadratic sum of the 
separate statistical and systematic errors. 
(Bottom)$95\%$ CL search reaches for $M_H$ 
as a functions of the LC integrated luminosity arising from the transverse 
polarization asymmetry in $e^+e^- \to e^+e^-$. From bottom to top the curves 
correspond to center of mass energies of 0.5, 0.8, 1, 1.2 and 1.5 TeV, 
respectively.}
\end{figure}

Let us begin the present analysis 
by considering the case of Bhabha scattering, \ie, $e^+e^-\to 
e^+e^-$. Extracting the overall electromagnetic coupling factors the helicity 
amplitudes are given by
\begin{eqnarray}
T_{+-}^{~+-} &=& -(1+z)\Big[1+{s\over {t}}+g_R^2({s\over {(s-M_Z^2)}}+{s\over 
{(t-M_Z^2)}})\Big]+f_g\Big[2{u\over {s}}+{3\over {4}}(1+{t\over {s}})
\Big]\, \nonumber \\
T_{-+}^{~-+} &=& -(1+z)\Big[1+{s\over {t}}+g_L^2({s\over {(s-M_Z^2)}}+{s\over 
{(t-M_Z^2)}})\Big]+f_g\Big[2{u\over {s}}+{3\over {4}}(1+{t\over {s}})
\Big]\, \nonumber \\
T_{+-}^{~-+} &=& T_{-+}^{~+-}=-(1-z)\Big[1+g_Rg_L{s\over {(s-M_Z^2)}}\Big]-f_g
\Big({3\over {4}}+{t\over {s}}\Big)\, \nonumber \\
T_{++}^{~++} &=& T_{--}^{~--}=-\Big[{s\over {t}}+g_Rg_L{s\over {(t-M_Z^2)}}
\Big]-f_g\Big(1+{3\over {4}}{t\over {s}}\Big)\,,
\end{eqnarray}
where $z=\cos \theta$, $t(u)=-s(1\mp z)/2$, $g_L=(-1/2+s_w^2)/(s_wc_w)$ and 
$g_R=s_w/c_w$ with $s_w^2=\sin ^2\theta_w \simeq 0.23$ being the 
conventional weak mixing 
angle. We observe that, unlike the case of the $f\bar f$ final state, there 
are non-annihilation channel amplitudes present, \ie, $T_{\pm\pm}$, 
corresponding to the $t-$ channel diagrams. Note that the conditions of 
Eq.(1) hold for this set of amplitudes. 
Note further that the spin-2 exchange merely augments the amplitudes which are 
already present in the SM (though with different $\cos \theta$ 
dependencies), \ie, no new helicity amplitudes are generated by spin-2 over 
those already due to spin-1. 

\vspace*{-0.5cm}
\nn
\begin{figure}[htbp]
\centerline{
\includegraphics[width=8cm,angle=90]{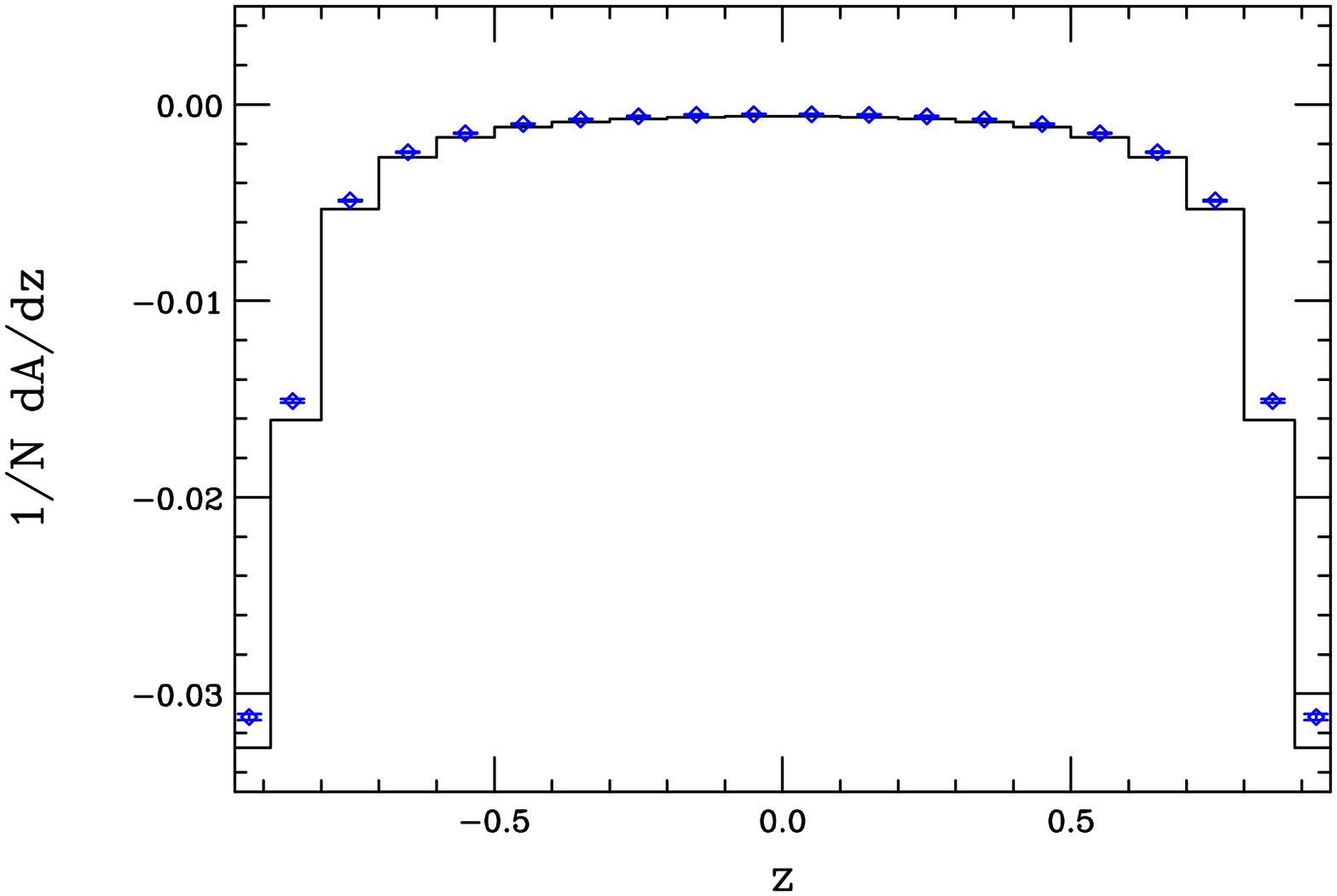}}
\vspace*{0.45cm}
\centerline{
\includegraphics[width=8cm,angle=90]{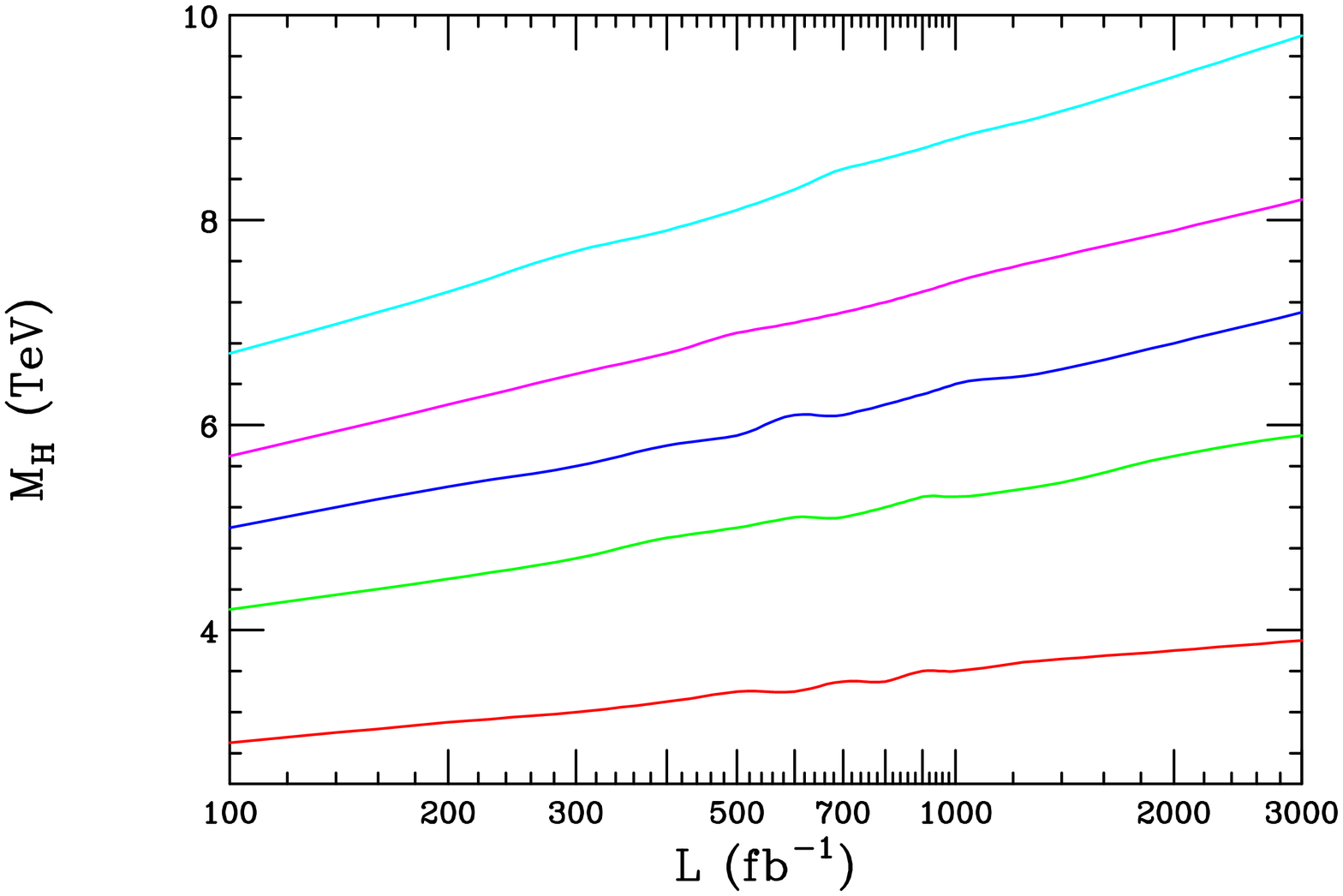}}
\vspace*{0.25cm}
\caption{Same as the previous figure but now for M{\o}ller scattering, $e^-e^-
\to e^-e^-$.}
\end{figure}

From the above amplitudes we can immediately obtain the bin-integrated 
TP asymmetry using Eqs. (2) and (3); a sample result is  shown in the top 
panel of Fig.~1 for the case of a $\sqrt s=500$ GeV LC assuming an integrated 
luminosity of 500 $fb^{-1}$. In this panel the SM prediction is compared 
to that obtained in the ADD model assuming $M_H=3\sqrt s=1.5$ TeV. Due to the 
existence of the $t-$channel QED pole, an angular cut $z\leq 0.95$ has been 
applied. In the SM, for $s>>M_Z^2$, the TP asymmetry scales{\cite {tp}} as 
$\sim -(1+z)^2$, which is a fair approximation to what we see here. 
For $z<0$ we see that the ADD prediction lies somewhat above the SM 
but drops significantly below it once large positive $z$ values are reached. 
It is clear from this figure that large search reaches are not likely in 
this channel, an expectation borne out by the results shown in the lower 
panel. (Recall that the contribution of the graviton exchange terms scale  
approximately as $M_H^{-4}$.) 
Here we see that the $95\%$ CL reach in the $\sqrt s=$500 GeV case is 
only about 2.5 TeV or $\sim 5\sqrt s$ with similar results holding for 
larger center of mass energies. Since it is likely that other forms of NP, 
such as a $Z'$,  
can also cause similar distortions in the shape of the TP asymmetry 
distribution, there is no unique graviton signature in this case.

The corresponding amplitudes for M{\o}ller scattering can be obtained by 
crossing and directly lead to the corresponding TP asymmetry results in this 
case as is shown in Fig.~2. Here a cut of $|z| \leq 0.95$ has been applied to 
remove the $u-$ and $t-$channel QED poles. In the central $z$ region the TP 
asymmetry is predicted to be very close to the SM in the ADD scenario 
differing only once $|z|\geq 0.5$ or so. Following the same procedures as 
in the case of Bhabha scattering we obtain the search reaches shown in  the 
lower panel of Fig.~2. Here we see that somewhat larger reaches are 
obtained, \ie, $7-8\sqrt s$, but these are still smaller than those obtained 
in our earlier work. The increased reach in M{\o}ller vs. Bhabha scattering 
using TP asymmetries is similar to what happens in the case of conventional  
contact interaction searches employing longitudinal polarization{\cite {pp}}. 
As in the case of Bhabha scattering there is no unique signature for graviton 
exchange in this process. 

\vspace*{-0.5cm}
\nn
\begin{figure}[htbp]
\centerline{
\includegraphics[width=8cm,angle=90]{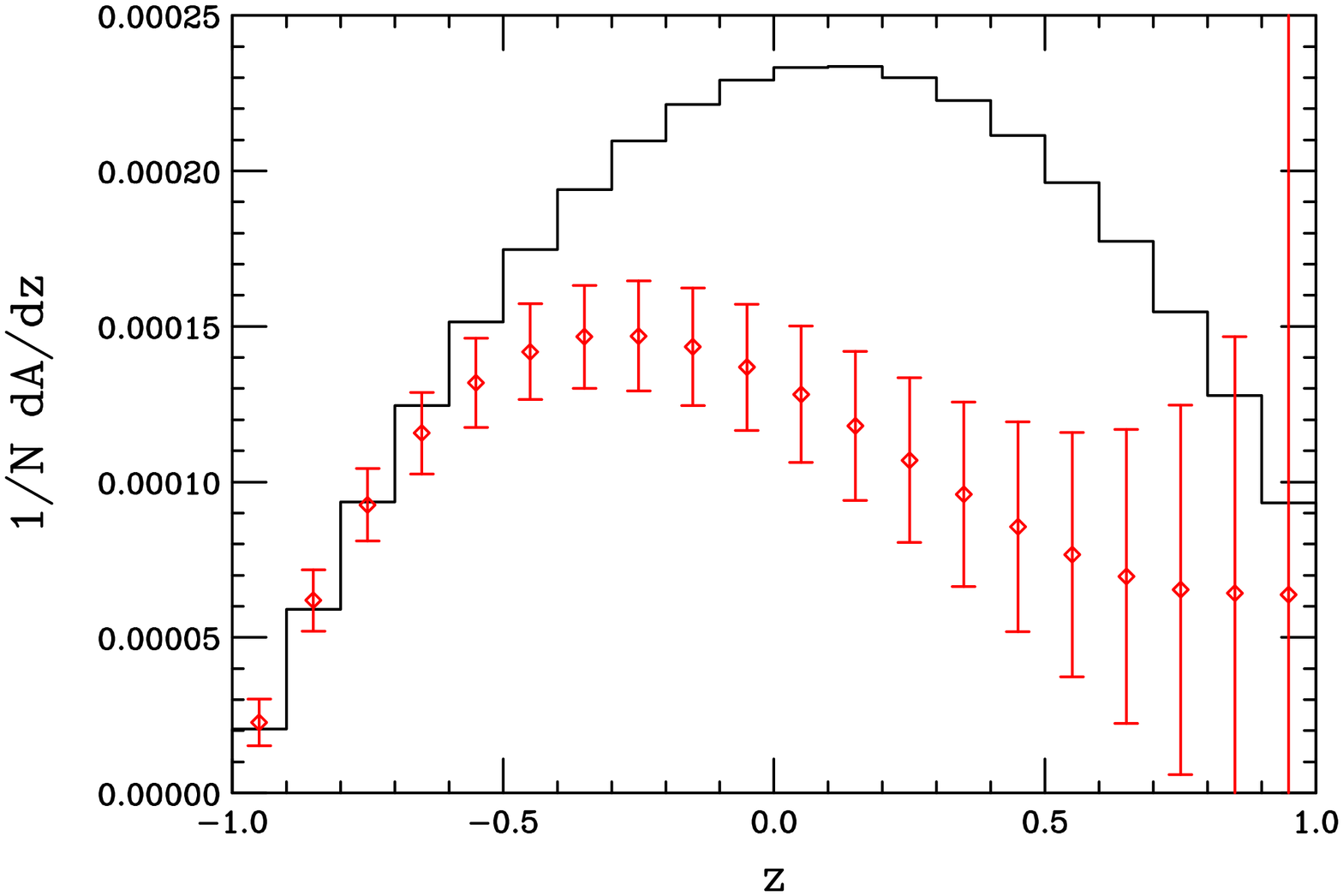}}
\vspace*{0.45cm}
\centerline{
\includegraphics[width=8cm,angle=90]{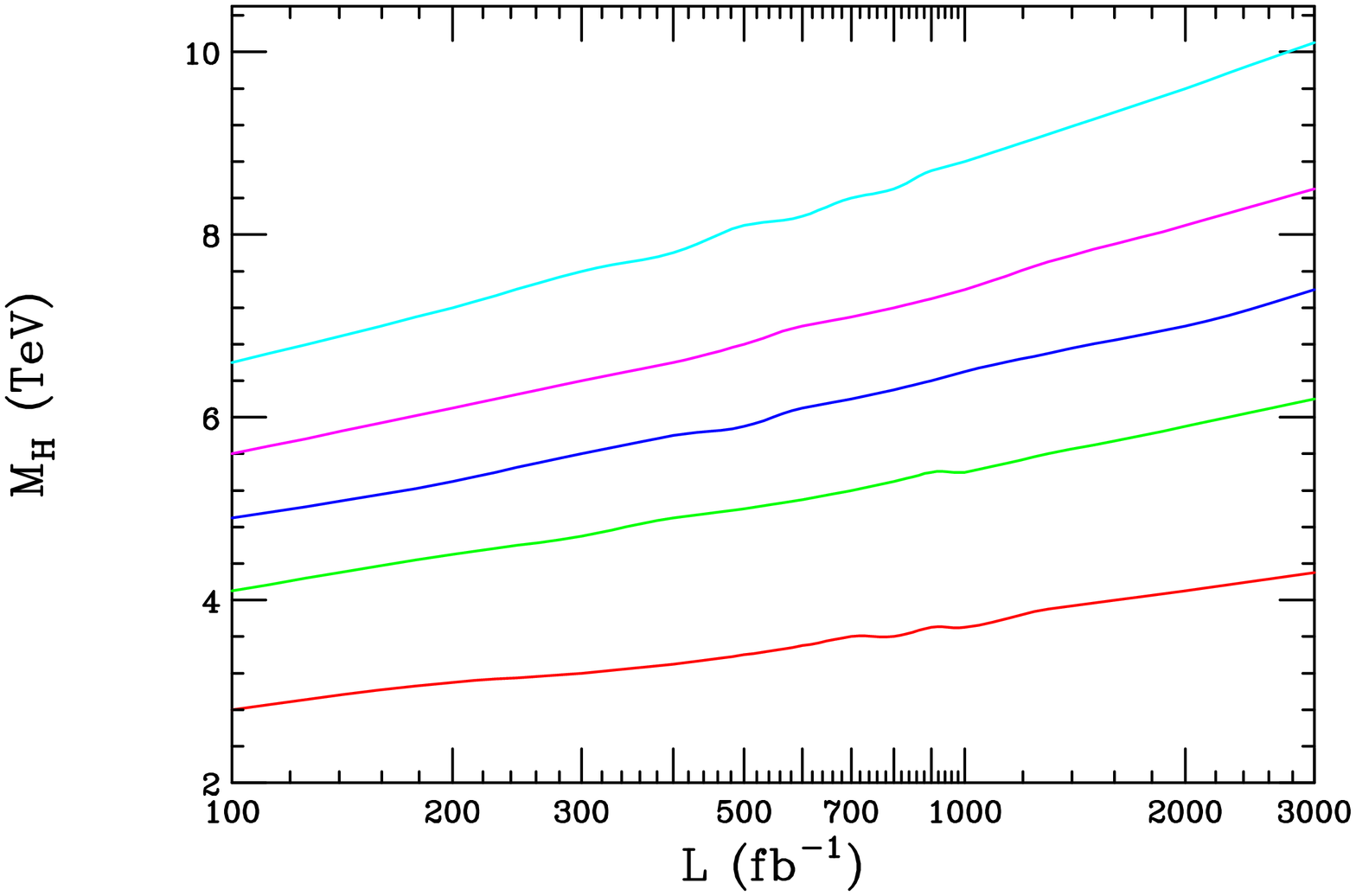}}
\caption{Same as the previous figure but now for the process $e^+e^- \to 
W^+W^-$.}
\vspace*{0.25cm}
\end{figure}

Let us now turn to the case of gauge boson pairs in the final state: 
$W^+W^-,2\gamma$ and $2Z$. For these cases our labor is relatively easy: 
all of the helicity amplitudes for these 
processes in the ADD model can be obtained directly from the work of Agashe 
and Deshpande{\cite{agashe}}. A quick analysis 
shows that the set of conditions analogous to Eq.(1) are satisfied for all 
of these final states. 

We consider the $W^+W^-$ final state first; the helicity amplitudes for 
this process are rather complicated and visually uninformative so we will 
not reproduce them here. A sample TP asymmetry in this 
case is shown in Fig. 3. Note that the SM shape is close to $1-z^2$ 
but deviates in detail from this in a $z$ asymmetric manner. 
In the ADD case, the TP 
asymmetry lies close to the SM in the backwards direction but 
falls significantly below it 
once $z>-0.5$ is reached. We also see that the distortion 
in the asymmetry due to 
graviton exchange is quite significant in the case where $M_H=1.5$ TeV.
As one might expect from this observation , the search reach for the ADD 
model in this case is somewhat larger than those obtained for the 
two processes above, $\simeq 8\sqrt s$ for a 500 GeV LC, as can be 
seen from the lower panel in Fig. 3. As in the cases above there is no 
obviously unique signature for spin-2/graviton exchange from this process.

\vspace*{-0.5cm}
\nn
\begin{figure}[htbp]
\centerline{
\includegraphics[width=8cm,angle=90]{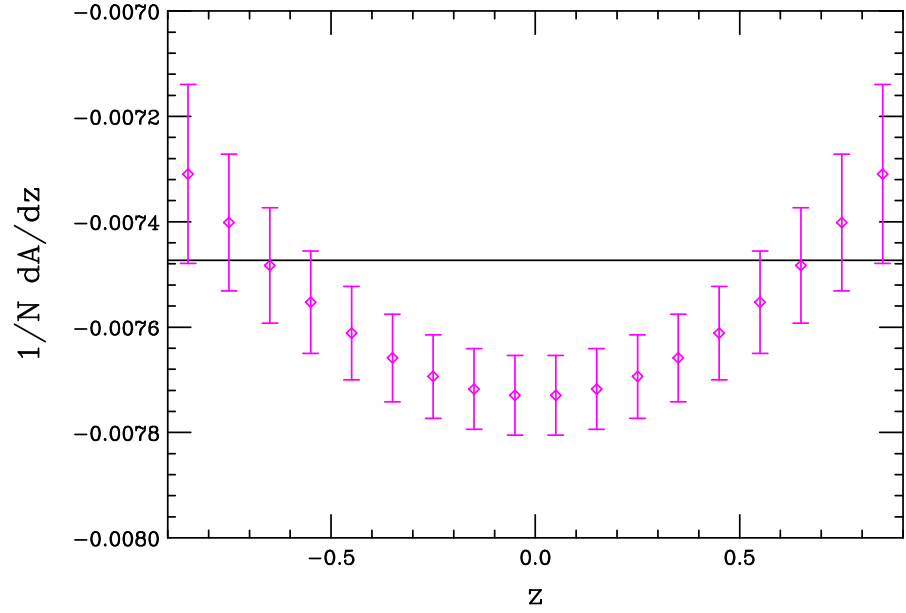}}
\vspace*{0.45cm}
\centerline{
\includegraphics[width=8cm,angle=90]{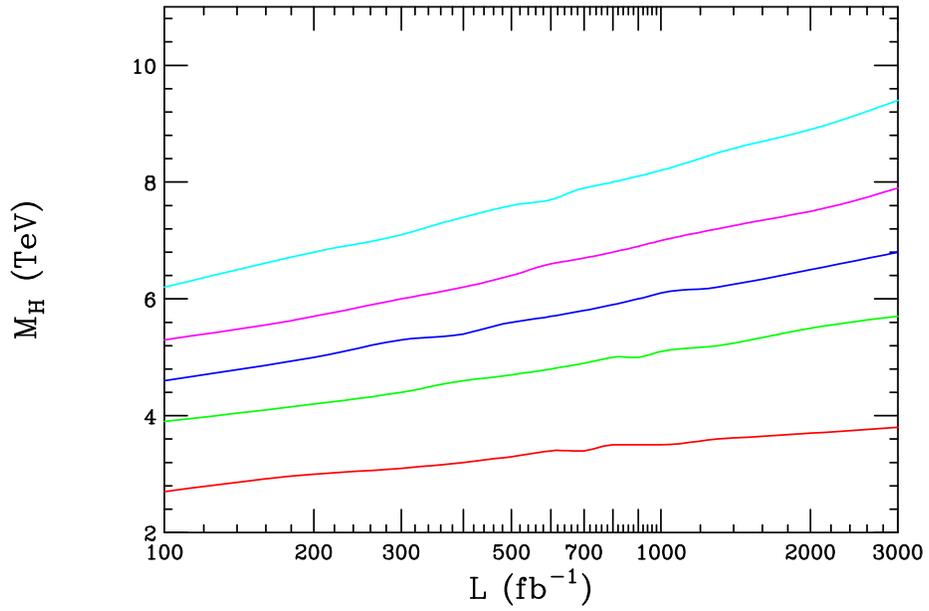}}
\vspace*{0.25cm}
\caption{Same as the previous figure but now for the process $e^+e^- \to 
2\gamma$.}
\end{figure}

The next possibility we consider is the pure QED 
process $e^+e^-\to 2\gamma$. In 
the SM the TP asymmetry is predicted to be $z$-independent as is shown in 
Fig. 4. Here, the existence of rather conventional sources of contact 
interactions, such as a $Z'$ or bilepton, cannot alter the SM predictions 
for this reaction. The lowest dimension operators that can contribute here are 
of the form $\sim \bar eeF_{\mu\nu}F^{\mu\nu}$, which might be induced by 
compositeness, and $\sim T_{\mu\nu}T^{\mu\nu}$, as can be induced by 
compositeness or graviton exchange. Clearly a unique signature for graviton 
exchange is not possible using this process. 
A short analysis shows that the ADD contributions simply 
adds to the SM TP=constant term by a relative factor of  
$\sim f_g (1-z^2)$ as can also seen in Fig.4. 
We can see this immediately from the helicity amplitudes for this 
process which are particularly simple:
\begin{eqnarray}
T_{+-}^{~+-}&=&-T_{-+}^{~-+}=-2{(1+z)\over {(1-z^2)^{1/2}}}
\Big(1-f_g{ut\over {s^2}}\Big)\, \nonumber \\
T_{-+}^{~+-}&=&-T_{+-}^{~-+}=-2{(1-z)\over {(1-z^2)^{1/2}}}
\Big(1-f_g{ut\over {s^2}}\Big)\,, 
\end{eqnarray}
where an overall factor of $e^2$ has been scaled out as before. What 
are the reaches 
for this process? Note that as in the calculations above we will 
employ an angular cut, $|z| \leq 0.95$, to remove the SM poles. Fig.4 shows 
the search reach for the ADD model using this TP asymmetry; it is quite 
respectable in comparison to the others we have found above $\sim 7\sqrt s$ 
for the case of a 500 GeV LC.

\vspace*{-0.5cm}
\nn
\begin{figure}[htbp]
\centerline{
\includegraphics[width=8cm,angle=90]{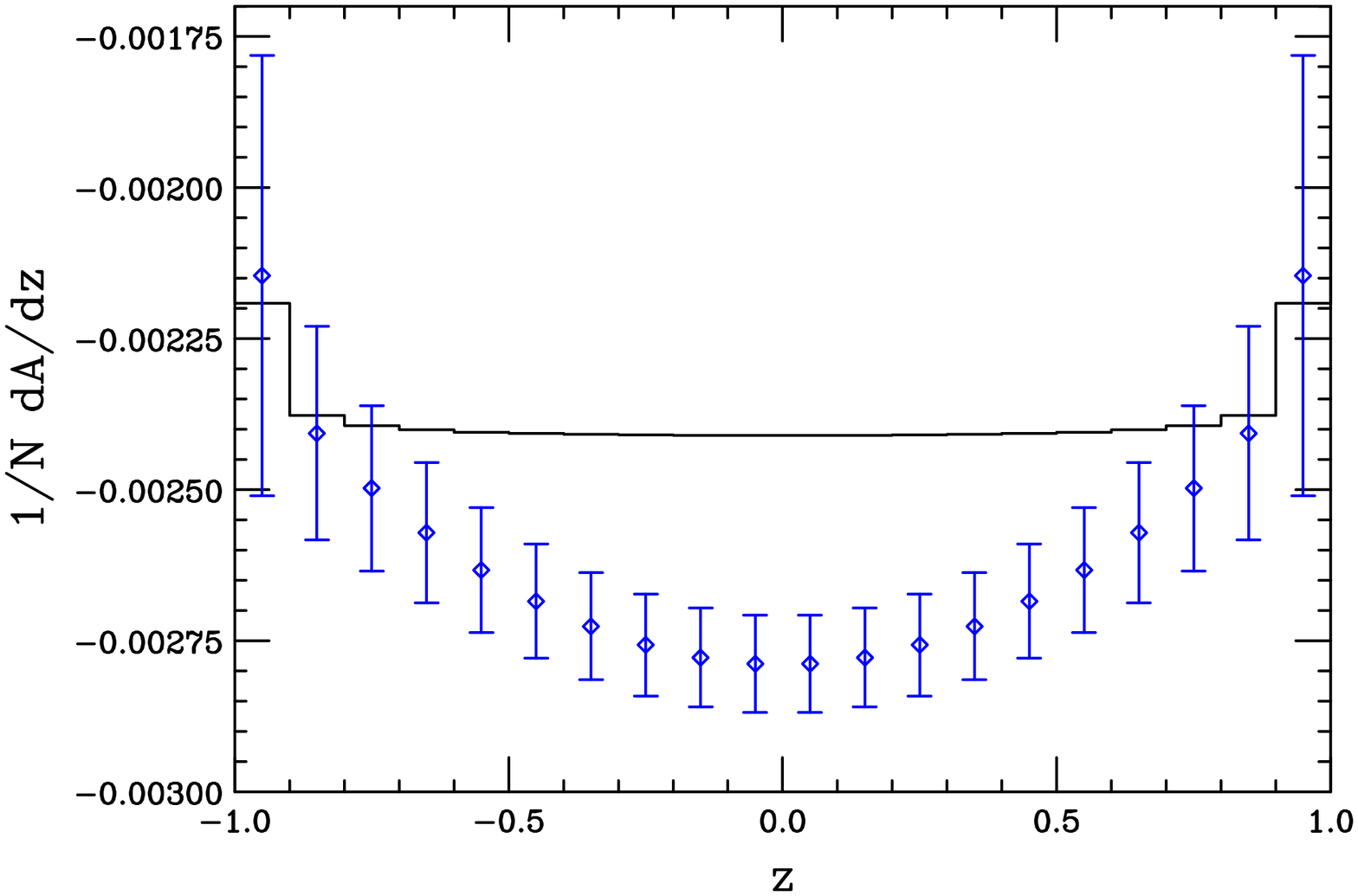}}
\vspace*{0.45cm}
\centerline{
\includegraphics[width=8cm,angle=90]{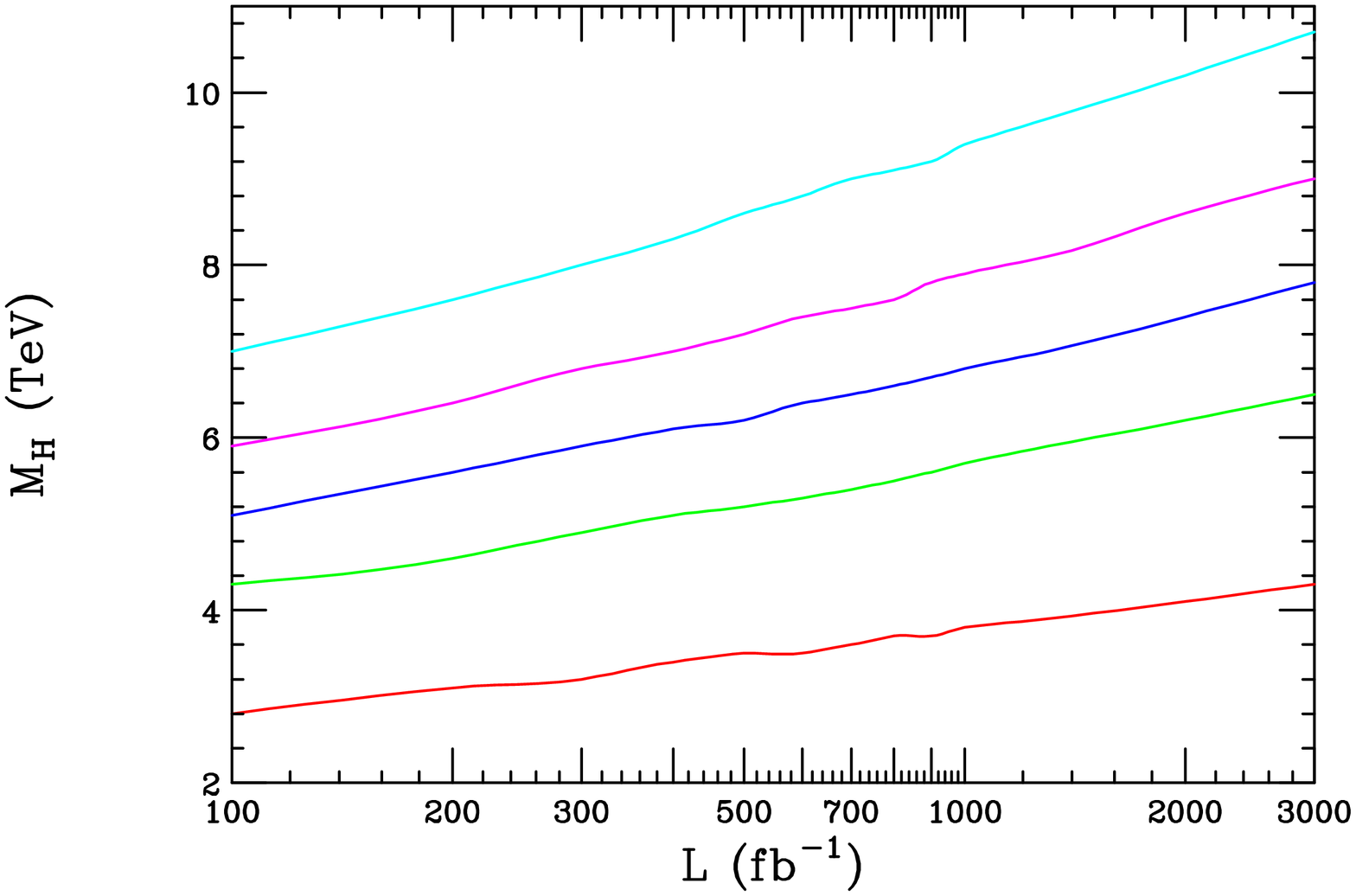}}
\vspace*{0.25cm}
\caption{Same as the previous figure but now for the process $e^+e^- \to 2Z$.}
\end{figure}

For the $2Z$ final state we would expect the TP asymmetry to 
behave similarly to the case of the $2\gamma$ final state apart from 
$M_Z^2/s$ corrections which are most visible near the would-be forward and 
backward poles. 
The SM $u-$ and $t-$ channel poles encountered in the $2\gamma$ final state 
case are thus smoothed out by the finite $Z$ boson mass so that no cuts are 
required. The helicity amplitudes for this process with the final states 
containing only transversely polarized $Z$'s are quite similar 
to those for the $2\gamma$ final state except for appropriate insertions of 
factors of $\beta=(1-4M_Z^2/s)^{1/2}$. Additional amplitudes corresponding to 
$Z$'s with longitudinal polarization are also present, however. 
As for the $2\gamma$ final state, we know that deviations in the observables 
associated with the $2Z$ final state can arise from a number of higher 
dimensional operators (including in this case anomalous triple gauge boson 
couplings) that we cannot uniquely trace back to graviton exchange. 
Fig. 5 shows that the TP asymmetry for the $2Z$ final state is nearly 
$z-$independent except in the very forward and backward directions in the SM. 
ADD graviton exchange, as in the $2\gamma$ case, induces a $\sim 1-z^2$-like 
contribution to the asymmetry as is also shown here. Since the rate for the 
$2Z$ and $2\gamma$ final states are similar(after cuts), as are the SM 
and ADD induced shapes of the TP polarizations, we might expect comparable 
search reaches. These are shown for the $2Z$ case in the lower panel of 
Fig. 5 and we see that our expectations are essentially confirmed. For a 
500 GeV LC a search reach of $\sim 7\sqrt s$ is obtained, essentially the 
same as for the $2\gamma$ final state.

\section{Discussion and Conclusions}

Disentangling the origin of contact interaction effects will require as many 
tools as possible. 
Transverse polarization asymmetries offer a special way to probe for NP in 
$e^+e^-$ processes and have been shown to be particularly sensitive to 
spin-2/graviton exchange for the case of the $f\bar f (f\neq e)$ set of 
final states. Not only does TP extend the conventional search reach but it 
also provides a means to uniquely identify spin-2 exchanges. Its utility for 
other final states has been, up to now, completely unknown. 

In this paper we have extended our previous analysis of TP asymmetries to 
encompass the processes $e^+e^-\to e^+e^-,W^+W^-,2\gamma$ and $2Z$ as well 
as $e^-e^-\to e^-e^-$ in order to access their sensitivity to graviton 
exchange within the context of the ADD model. The results of our analysis 
are twofold: ($i$) We have found that the various processes above lead to 
search reaches for the ADD cutoff scale, $M_H$, in the range of $(5-8)\sqrt s$ 
for a 500 GeV LC. These are very respectable reaches given that they are 
based on only a single observable and result from only a single final 
state. By contrast, the usual analyses which employ the total cross sections, 
angular distributions, tau polarization and the $A_{LR}'s$ for the various 
final states $f$ when combined have reaches that are only $\sim 10\sqrt s$. 
However, none of the final states we have studied here have search reaches 
as large as that obtained from the combined $f\bar f$ analysis. ($ii$) Our 
earlier analysis demonstrated that essentially only graviton exchange could 
shift the $1-z^2$ shape of the TP asymmetry distribution. Though 
the $f\bar f$ final states asymmetries 
are uniquely modified by the presence of 
spin-2/graviton exchange, this is no longer true for any of the processes 
we have examined here. Clearly more detailed studies are needed to verify 
these results. 

Hopefully signs of new physics will be observed soon after the turn on of 
future colliders.

\noindent{\Large\bf Acknowledgements}

The author would like to thank J.L. Hewett, G. Moortgat-Pick, K. Desch, 
J. Clendenin and M. Woods for discussion related to this work.

%
\def\MPL #1 #2 #3 {Mod. Phys. Lett. {\bf#1},\ #2 (#3)}
\def\NPB #1 #2 #3 {Nucl. Phys. {\bf#1},\ #2 (#3)}
\def\PLB #1 #2 #3 {Phys. Lett. {\bf#1},\ #2 (#3)}
\def\PR #1 #2 #3 {Phys. Rep. {\bf#1},\ #2 (#3)}
\def\PRD #1 #2 #3 {Phys. Rev. {\bf#1},\ #2 (#3)}
\def\PRL #1 #2 #3 {Phys. Rev. Lett. {\bf#1},\ #2 (#3)}
\def\RMP #1 #2 #3 {Rev. Mod. Phys. {\bf#1},\ #2 (#3)}
\def\NIM #1 #2 #3 {Nuc. Inst. Meth. {\bf#1},\ #2 (#3)}
\def\ZPC #1 #2 #3 {Z. Phys. {\bf#1},\ #2 (#3)}
\def\EJPC #1 #2 #3 {E. Phys. J. {\bf#1},\ #2 (#3)}
\def\IJMP #1 #2 #3 {Int. J. Mod. Phys. {\bf#1},\ #2 (#3)}
\def\JHEP #1 #2 #3 {J. High En. Phys. {\bf#1},\ #2 (#3)}

\end{document}